\documentclass[conference]{IEEEtran}
\IEEEoverridecommandlockouts
\usepackage{cite}
\usepackage{amsmath,amssymb,amsfonts}
\usepackage{algorithmic}
\usepackage{graphicx}
\usepackage{textcomp}
\usepackage{xcolor}
\def\BibTeX{{\rm B\kern-.05em{\sc i\kern-.025em b}\kern-.08em
    T\kern-.1667em\lower.7ex\hbox{E}\kern-.125emX}}

\usepackage[utf8]{inputenc} 
\usepackage[T1]{fontenc}    
\usepackage{hyperref}       
\usepackage{url}            
\usepackage{booktabs}       
\usepackage{amsfonts}       
\usepackage{nicefrac}       
\usepackage{microtype}      
\usepackage{lipsum}		
\usepackage{graphicx}
\usepackage{doi}
\usepackage{braket}
\usepackage{subcaption}

\begin{document}

\title{Machine Learning based Discrimination for \\Excited State Promoted Readout}

\author{\IEEEauthorblockN{Utkarsh Azad}
\IEEEauthorblockA{\textit{Center for Computational Natural Sciences and Bioinformatics}, \\
\textit{Center for Quantum Science and Technology} \\
\textit{International Institute of Information Technology}\\
Hyderabad, TS, India\\
utkarsh.azad@research.iiit.ac.in}
\and
\IEEEauthorblockN{Helena Zhang}
\IEEEauthorblockA{\textit{IBM Quantum} \\
\textit{IBM Thomas J. Watson Research Center} \\
Yorktown Heights, NY, USA \\
}
}

\maketitle

\begin{abstract}
A limiting factor for readout fidelity for superconducting qubits is the relaxation of the qubit to the ground state before the time needed for the resonator to reach its final target state. A technique known as excited state promoted (ESP) readout was proposed to reduce this effect and further improve the readout contrast on superconducting hardware. In this work, we use readout data from IBM's five-qubit quantum systems to measure the effectiveness of using deep neural networks, like feedforward neural networks, and various classification algorithms, like k-nearest neighbors, decision trees, and Gaussian naive Bayes, for single-qubit and multi-qubit discrimination. These methods were compared to standardly used linear and quadratic discriminant analysis algorithms based on their qubit-state-assignment fidelity performance, robustness to readout crosstalk, and training time.
\end{abstract}


\section{Introduction}
Quantum computers are speculated to have a computational edge over their classical counterparts in solving problems with better accuracy and lesser computational time in various areas, such as quantum chemistry \cite{doi:10.1021/acs.chemrev.8b00803}, molecular simulations \cite{Yuan2019theoryofvariational}, machine learning \cite{10.1038/s41567-019-0648-8}, etc. Even though there has been a recent report of quantum advantage by Xanadu \cite{10.1038/s41586-022-04725-x}, and previously by Google \cite{10.1038/s41586-019-1666-5} which has been usurped recently \cite{10.1103/PhysRevLett.129.090502}, in reality, quantum hardware requires millions of good quality qubits to achieve such an advantage in the abovementioned areas for doing something useful. But, the present-day hardware, generally called noisy intermediate-scale quantum (NISQ) hardware, contains not more than a couple of hundred qubits that are error-prone, severely limiting their computational capabilities. The major problem in scaling up these devices is having efficient qubit control and high readout fidelity. Hence, considerable work is required to retain and further improve these systems if we want to increase their size and complexity, particularly for combating errors at all stages of the computational pipeline: initialization, execution, and readout.

In this work, we present machine-learning-enabled qubit-state discrimination utilizing excite-state-promoted (ESP) readout, which is a way to improve qubit readout fidelity in a scalable way by using extra levels of transmon qubits, i.e., exciting the $\ket{1}$ state to the $\ket{2}$ state for readout \cite{2009NatPhMallet}. This essentially changes the discrimination problem from a two-state (Fig. \ref{fig:two-state}) system to a three-state (Fig. \ref{fig:three-state}) system. We evaluate the qubit-state discrimination performance of various machine learning models such as k-nearest neighbors (KNN) \cite{Altman_1992}, decision trees \cite{Rokach_2013}, Gaussian naive Bayes (GNB) \cite{10.5555/1671238}, linear and quadratic discriminant analysis (LDA and QDA) \cite{Tharwat_2016}, and a fully-connected neural network (FNN). To evaluate these different qubit-state discriminator techniques, we use ESP readout outputs from five of the IBM's five-qubit quantum systems \cite{IBMQ}: (i) \textit{ibmq\_rome}, (ii) \textit{ibmq\_bogota}, (iii) \textit{ibmq\_merlin}, (iv) \textit{ibmq\_belem}, and (v) \textit{ibmq\_quito}. We examine their qubit-state assignment performance using a confusion matrix and the cross-fidelity metric introduced in \cite{10.1103/physrevapplied.17.014024}. We show that classifiers based on FNN and GNB outperform LDA and QDA in both single- and multi-qubit discrimination tasks. 

\section{\label{sec:sec-2}Theory of qubit readout}

Over the past two decades, superconducting qubits have emerged as a leading quantum computing platform that has been pursued by various leading industries such as IBM \cite{IBMQ}, Google \cite{comp_cirq}, Rigetti \cite{ccquad_Pyquil}, etc. The basis of their quantum hardware is a particular kind of superconducting qubit, popularly known as the transmon qubit, which is composed of a Josephson junction and capacitor \cite{2019ApPRvK}. The Hamiltonian of the transmon can be described as:

\begin{figure*}[!th]
    \centering
    \begin{subfigure}[b]{.40\linewidth}
    \begin{minipage}{.1\textwidth}
        \caption{}
        \label{fig:qho}
    \end{minipage}%
    \begin{minipage}{0.95\textwidth}
        \includegraphics[width=.9\linewidth]{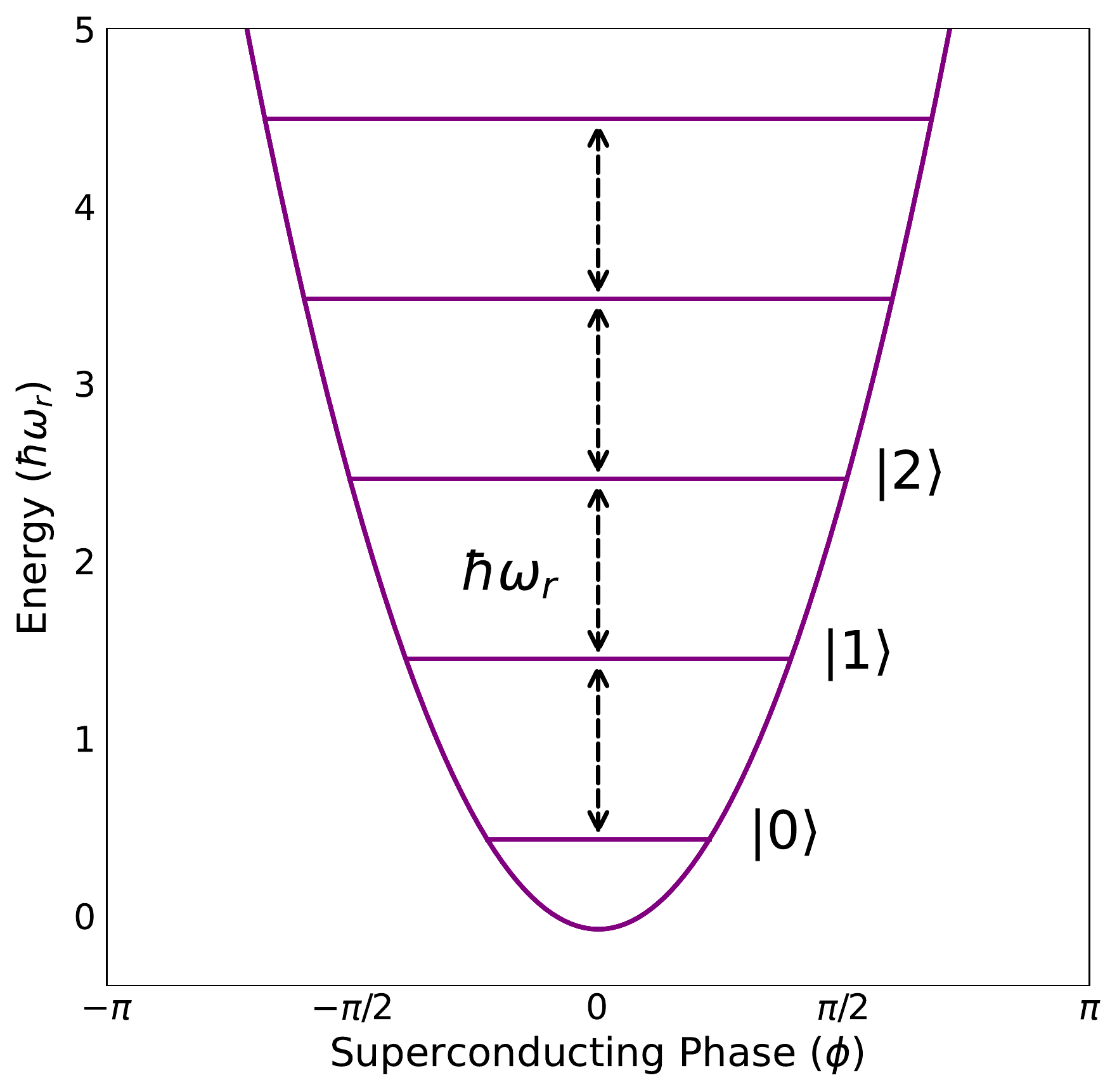}
    \end{minipage}
    \end{subfigure}
    \begin{subfigure}[b]{.40\linewidth}
    \begin{minipage}{.1\textwidth}
        \caption{}
        \label{fig:transmon}
    \end{minipage}%
    \begin{minipage}{0.95\textwidth}
        \includegraphics[width=.9\linewidth]{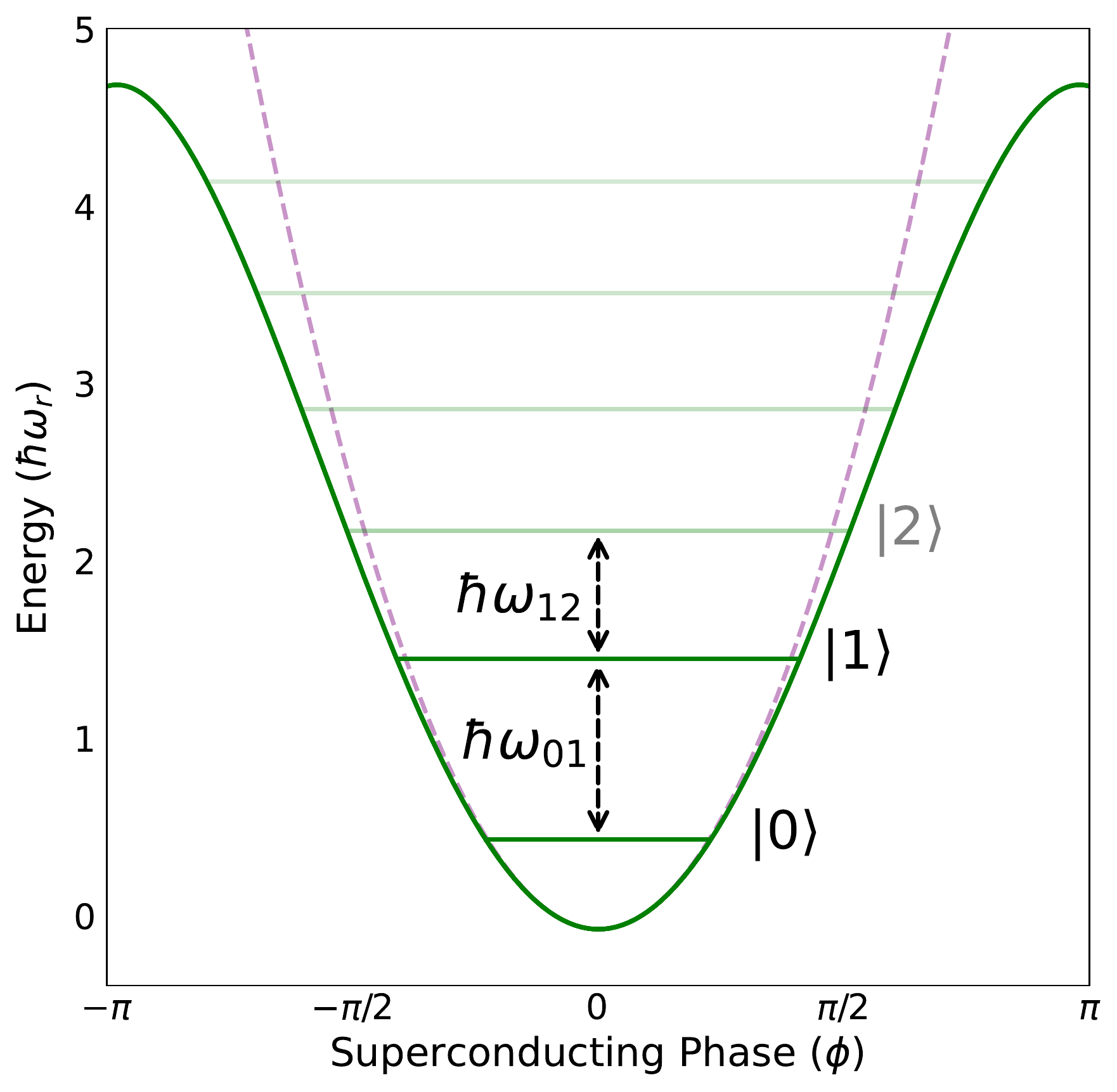}
    \end{minipage}
    \end{subfigure}
    \caption{(a) Energy levels are equidistantly spaced by $\hbar\omega_r$ in the energy potential for the quantum harmonic oscillator (QHO). (b) The inclusion of Josephson inductance leads to non-equidistant energy levels as it changes the quadratic energy potential (dashed purple) into sinusoidal potential (solid green), i.e., by including higher order terms. The sufficient difference in the energy spacing $\hbar\omega_{01}$ and $\hbar\omega_{12}$ allows us to form a computational basis by isolating the two lowest energy levels $\ket{0}$ and $\ket{1}$ of the transmon. Both figures here have been adapted from \cite{2019ApPRvK}.} 
    \label{fig:ep}
\end{figure*}
\begin{equation}\label{eq:eq1}
    H = 4 E_C n^2 - E_J \cos(\phi),
\end{equation}
where $E_C, E_J$ denote the energies of the capacitor and Josephson junction present in the superconducting circuit, $n$ is the reduced charge number operator, and $\phi$ is the reduced flux across the Josephson junction with $\hbar=1$. 

\subsection{\label{subsec:sec-2-1}Transmon qubits}

In principle, the variable $\phi$ can acquire a range of values, but the system begins to behave as a transmon qubit only in the regime where $\phi \rightarrow 0$. This allows us to approximate H (\ref{eq:eq1}) by performing a Taylor expansion of the $E_J\cos(\phi)$ while ignoring constant terms:
\begin{equation}
    \lim_{\phi \to 0} E_J \cos(\phi) \approx \frac{1}{2!} E_J \phi^2 - \frac{1}{4!} E_J \phi^4 + \mathcal{O}(\phi^6),
\end{equation}
where the quadratic term $\phi^2$ defines the standard quantum harmonic oscillator (\ref{fig:qho}), and the following subsequent higher-order terms contribute to anharmonicity in the system. This is important for the system to isolate the two lowest energy levels $\ket{0}$ and $\ket{1}$ and determine a computational basis (\ref{fig:transmon}), which would not be possible in the case of the standard quantum harmonic oscillator due to presence of energy levels which are equidistant.

It can be further shown that this system resembles a Duffing oscillator with the Hamiltonian $H_D$ and $n \sim (a-a^\dagger)$, $\phi \sim (a+a^\dagger)$ as the canonical conjugate variables, where $a$ ($a^{\dagger}$) is the annihilation (creation) operator of the qubit system \cite{2019ApPRvK}:
\begin{equation}
    H = \omega a^\dagger a + \frac{\alpha}{2} a^\dagger a^\dagger a a.
\end{equation}
Here, $\omega$ corresponds to the $\omega_{01}$, i.e., the excitation frequency from the ground state to the first excited energy state ($0\rightarrow1$), and $\alpha$ is the anharmonicity between the excitation frequencies $\omega_{01}$ and $\omega_{12}$. By tuning the $|\alpha|$ to sufficiently large values, one can isolate the standard two-dimensional subspace by suppressing leakage to the higher energy states (\ref{fig:ep}).

\begin{figure*}[!htp]
    \centering
    \begin{subfigure}[b]{\linewidth}
    \begin{minipage}{.1\textwidth}
        \caption{}
        \label{fig:two-state}
    \end{minipage}%
    \begin{minipage}{0.9\textwidth}
    \begin{subfigure}[b]{.3\linewidth}
    \begin{minipage}{0.92\textwidth}
        \includegraphics[width=\linewidth]{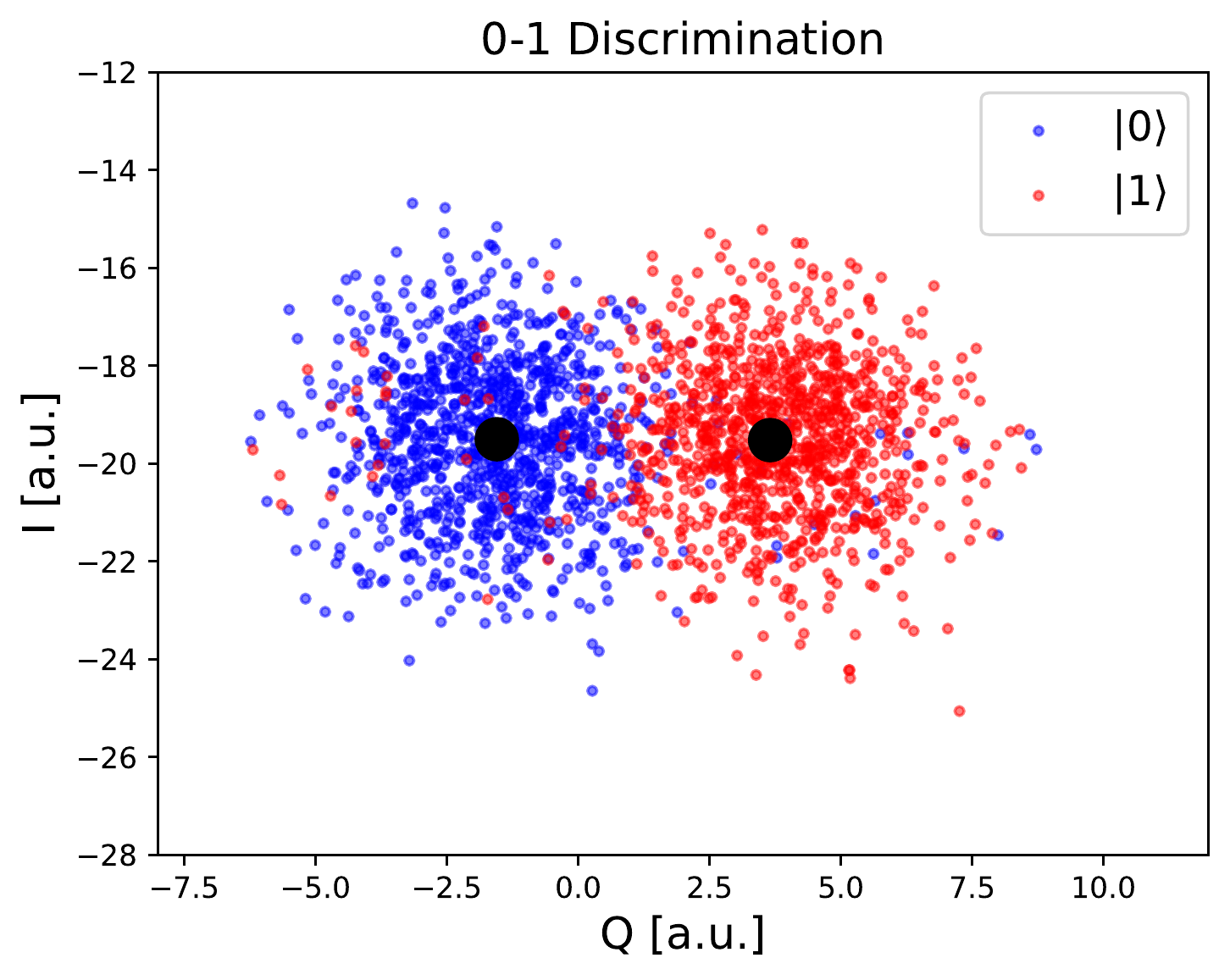}
    \end{minipage}
    \end{subfigure}
    \begin{subfigure}[b]{.3\linewidth}
    \begin{minipage}{0.92\textwidth}
        \includegraphics[width=\linewidth]{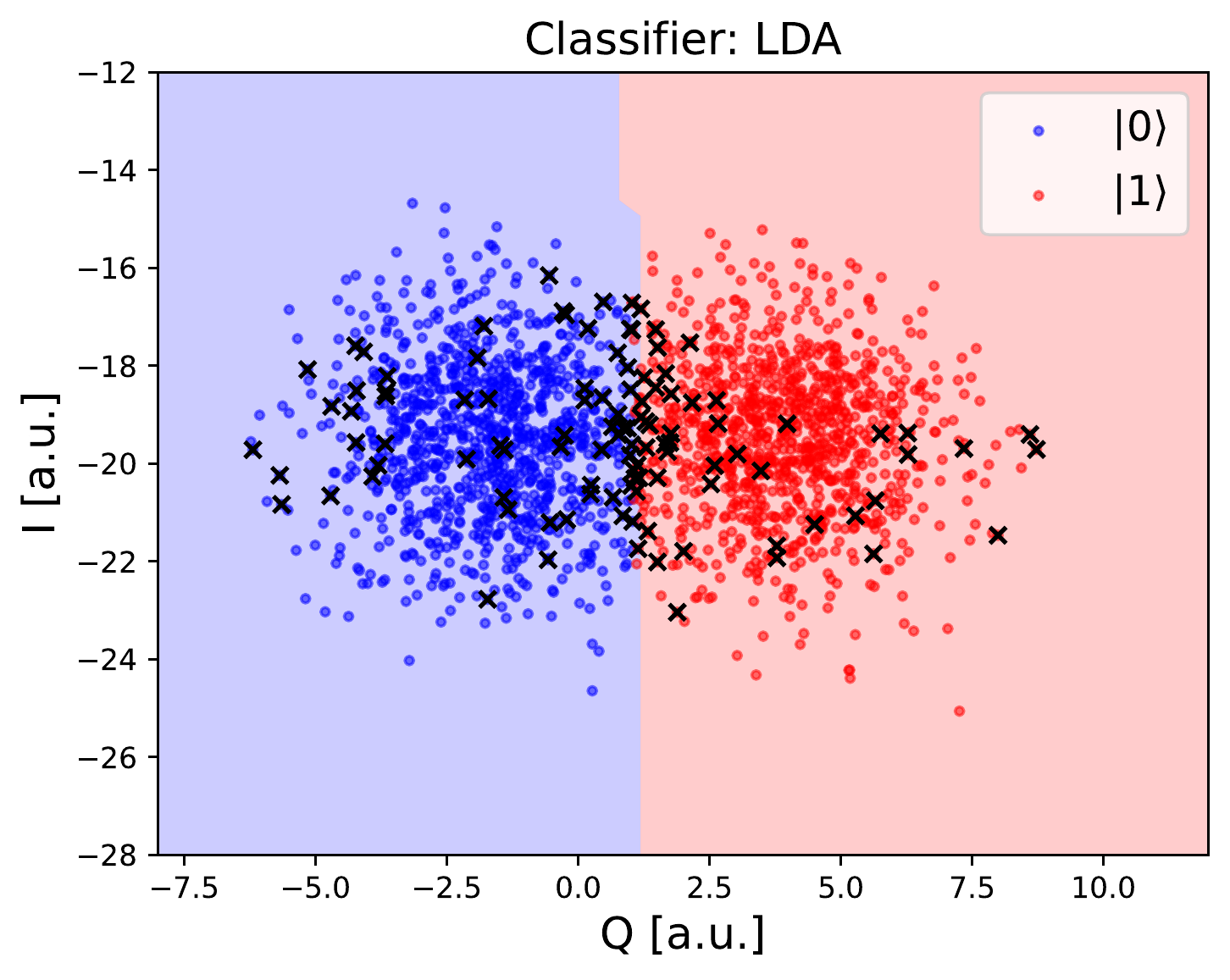}
    \end{minipage}
    \end{subfigure}
    \begin{subfigure}[b]{.3\linewidth}
    \begin{minipage}{0.92\textwidth}
        \includegraphics[width=\linewidth]{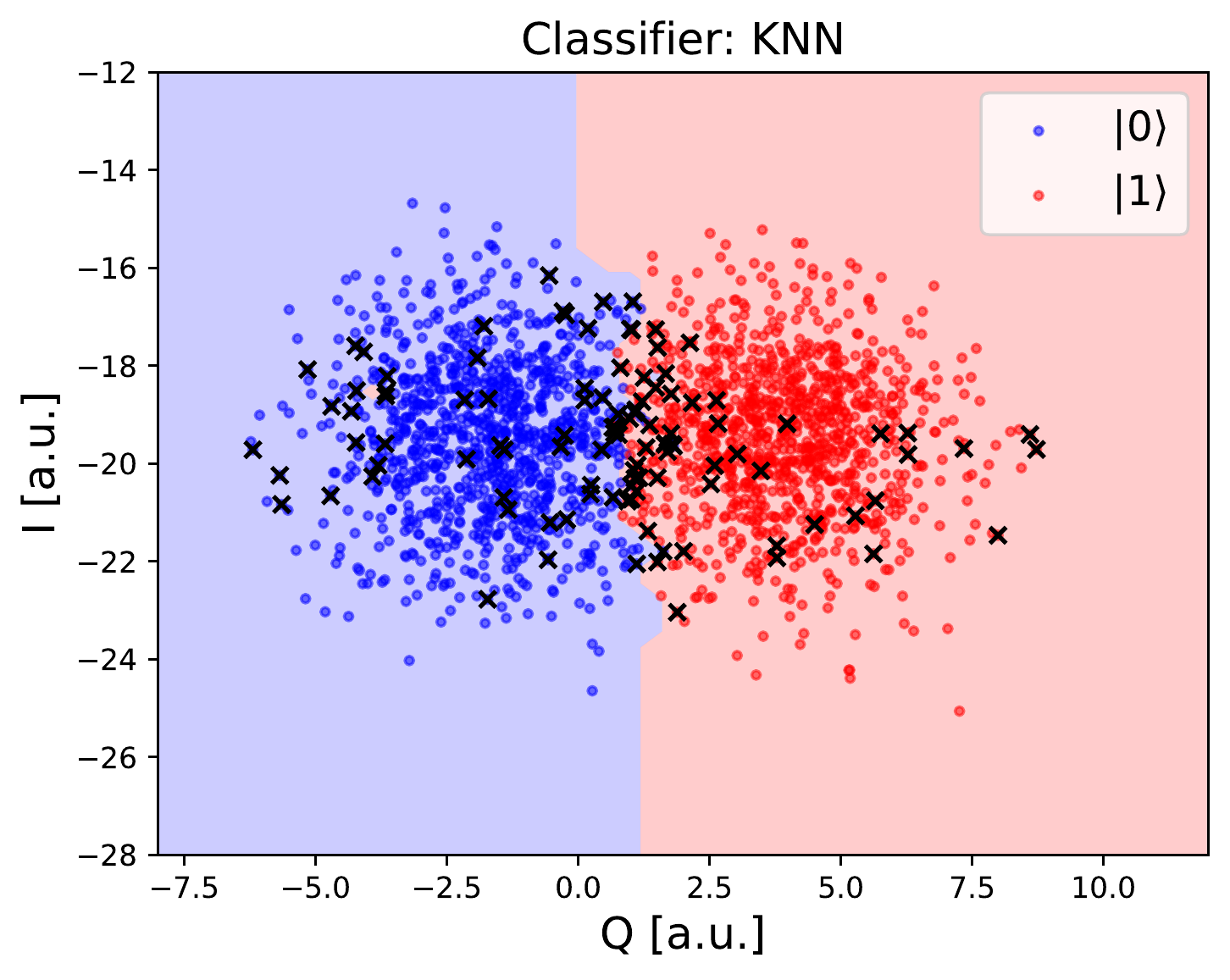}
    \end{minipage}
    \end{subfigure}
   
    \end{minipage}
    \end{subfigure}\\
    \begin{subfigure}[b]{\linewidth}
    \begin{minipage}{.1\textwidth}
        \caption{}
        \label{fig:three-state}
    \end{minipage}%
    \begin{minipage}{0.9\textwidth}
   \begin{subfigure}[b]{.3\linewidth}
    \begin{minipage}{0.92\textwidth}
        \includegraphics[width=\linewidth]{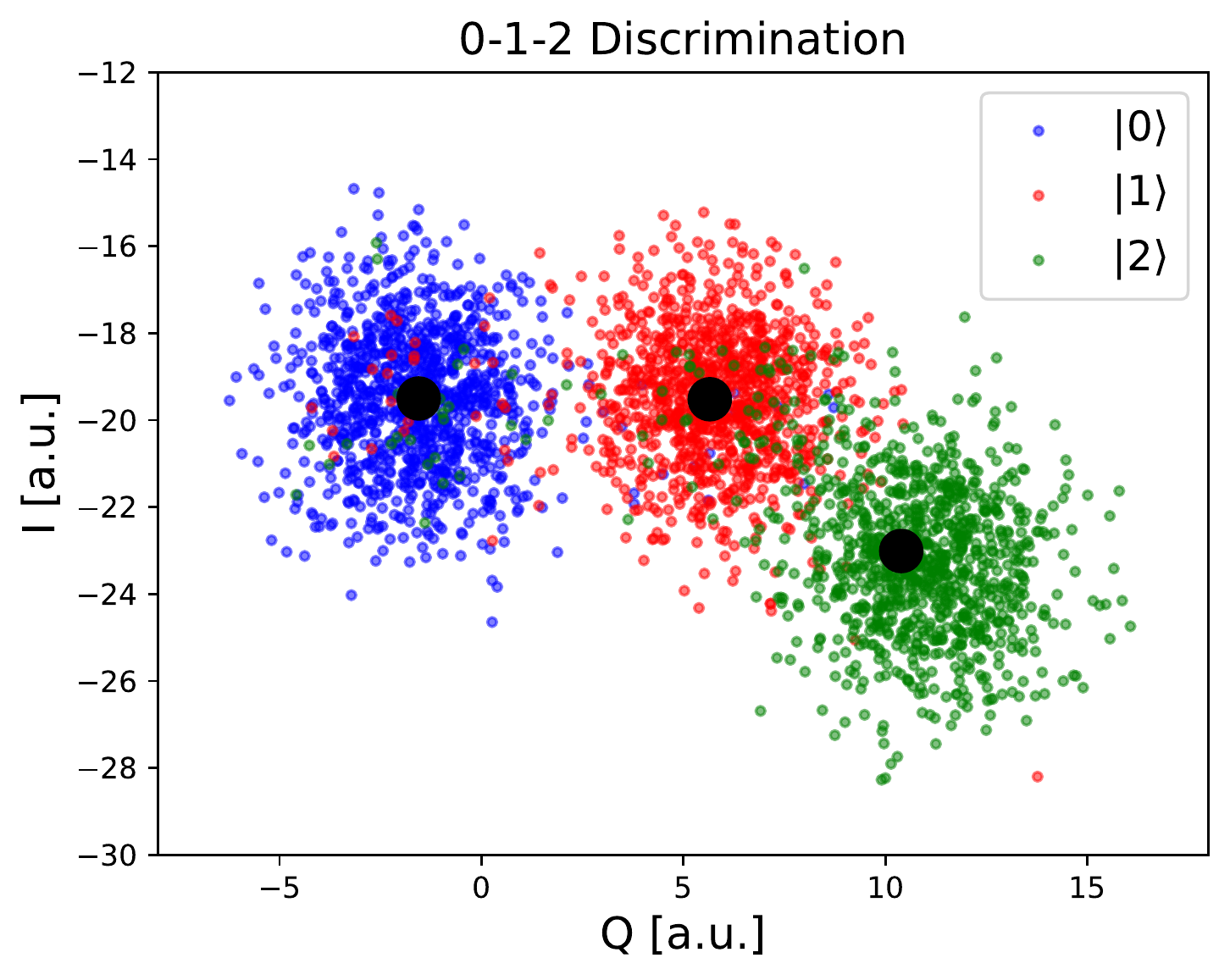}
    \end{minipage}
    \end{subfigure}
    \begin{subfigure}[b]{.3\linewidth}
    \begin{minipage}{0.92\textwidth}
        \includegraphics[width=\linewidth]{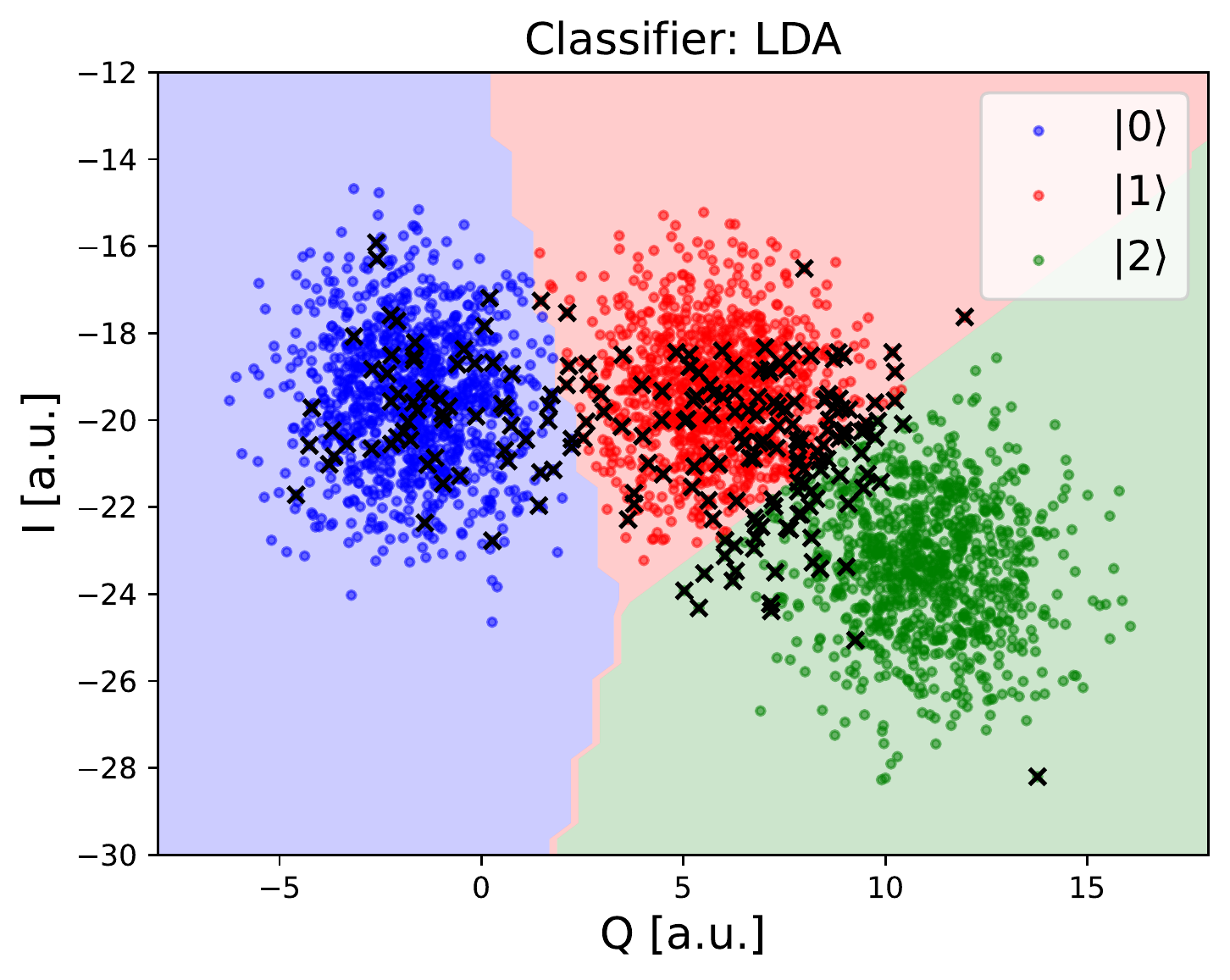}
    \end{minipage}
    \end{subfigure}
    \begin{subfigure}[b]{.3\linewidth}
    \begin{minipage}{0.92\textwidth}
        \includegraphics[width=\linewidth]{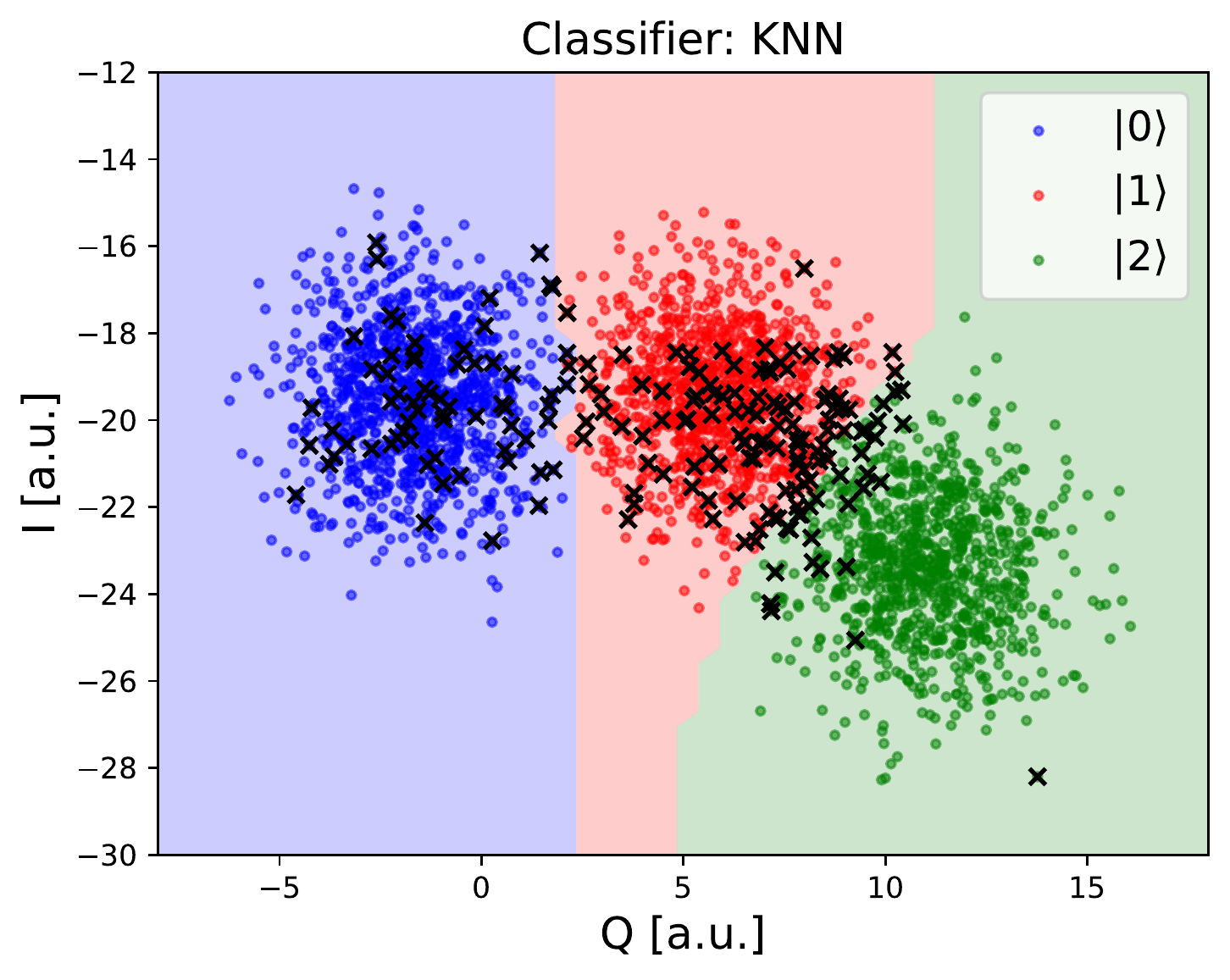}
    \end{minipage}
    \end{subfigure}
    \end{minipage}
    \end{subfigure}
    \caption{IQ plots for qubit states: (a) $\{\ket{0}, \ket{1}\}$ and (b) $\{\ket{0}, \ket{1}, \ket{2}\}$ are shown in the first column. In the other two columns, we show decision surfaces for their discrimination done via: (i) linear discriminant analysis (LDA) and (ii) k-nearest neighbor (KNN), where misclassified IQ points are shown as black crosses. The data presented here was acquired from \textit{ibmq\_armonk}.} 
    \label{fig:IQ-plots}
\end{figure*}

\subsection{\label{subsec:sec-2-2}Dispersive Readout}
The ability to perform high fidelity readout of the qubit states is a crucial cornerstone of any quantum processor. The most common technique utilized in the circuit QED architecture is that of dispersive readout. In this method, the qubit (quantum system of interest) is entangled with an observable of a superconducting resonator (probe), allowing us to gain information about the qubit state by interrogating the resonator - rather than directly interacting with the qubit. Therefore, readout performance depends on the signal-to-noise ratio of a microwave pulse tone sent to the resonator while minimizing the unwanted back-action on the qubit. 

For the purpose of this work, we can skip the details of the experimental implementation of qubit-state measurements and instead focus on the readout event itself. It commences with a short microwave tone directed to the resonator at the resonator probe frequency $\omega_{RO}$ (the carrier frequency), which acquires the following form after interacting with the resonator:
\begin{equation}
	s(t) = A_{RO} \cos(\omega_{RO}t + \theta_{RO}) = \mathbb{R} \Big\{ A_{RO} e^{j (\omega_{RO}t + \theta_{RO})} \Big\}, 
\label{eq:rsignal}
\end{equation}
where $A_{RO}$ and $\theta_{RO}$ are the qubit-state-dependent amplitude and phase that we wish to measure, and $\mathbb{R}$ represents the \textit{real} part of an expression. We can rewrite equation \ref{eq:rsignal} in the following \textit{phasor} notation form:
\begin{equation}\label{eq:phasor}
	s(t) = Re \Bigg\{ \underbrace{A_{RO} e^{(\theta_{RO})}}_{phasor} e^{j(w_{RO}t)} \Bigg\}.
\end{equation}
From equation \ref{eq:phasor}, we gather that performing qubit readout should be equivalent to (i) measuring the “in-phase” component $I$ and a “quadrature” component $Q$ of the complex number represented by the phasor, and (ii) determining the amplitude $A_{RO}$ and the phase $\theta_{RO}$:
\begin{equation}\label{eq:iq-data}
	A_{RO} e^{(\theta_{RO})} = A_{RO} \cos(\theta_{RO}) + j A_{RO} \sin(\theta_{RO}) \equiv I + jQ.
\end{equation}

\subsection{\label{subsec:sec-2-3}Discrimination}
As noted in the previous subsection, we extract the $I$ and $Q$ components from the readout signal for performing qubit readout. These components construct an $I-Q$ plane, as shown in Fig. \ref{fig:IQ-plots}. It is seen that the $(I,\ Q)$ valued coordinates form specific clusters on this plane, corresponding to the actual state of the qubit, and hence implying it to be of a particular energy level, $\ket{k}$. For example, in Fig. \ref{fig:two-state}, we see that two clusters marked by blue and red points correspond to the qubit states $\ket{0}$ and $\ket{1}$ states, respectively. Given this $IQ$ data on the plane, we incorporate a discriminator (or a classifier) to find the boundaries of the cluster formed by each state so that for the subsequent incoming $IQ$ output, we can predict the corresponding unknown state with sufficient confidence.

\begin{figure*}[!tp]
    \centering
    \begin{subfigure}[b]{.3\linewidth}
    \begin{minipage}{.1\textwidth}
        \caption{}
        \label{fig:l-top}
    \end{minipage}%
    \begin{minipage}{0.65\textwidth}
        \includegraphics[width=\linewidth]{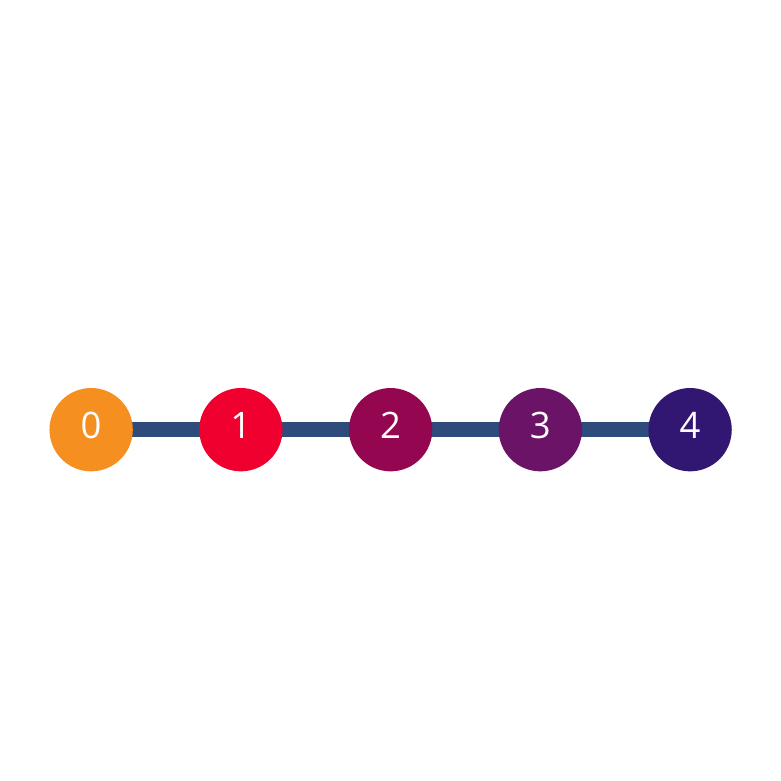}
    \end{minipage}
    \end{subfigure}
    \begin{subfigure}[b]{.3\linewidth}
    \begin{minipage}{.1\textwidth}
        \caption{}
        \label{fig:t-topo}
    \end{minipage}%
    \begin{minipage}{0.65\textwidth}
        \includegraphics[width=\linewidth]{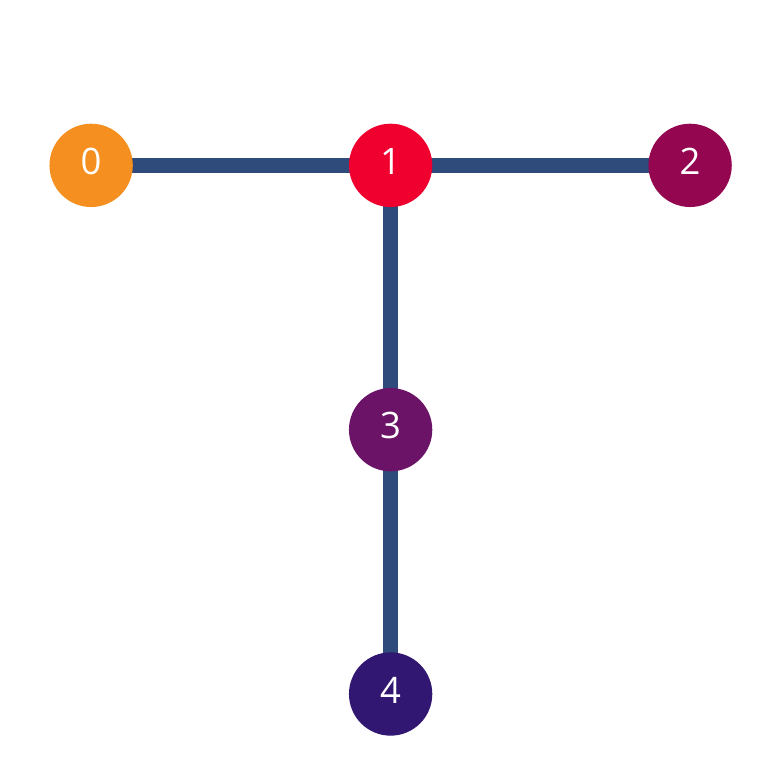}
    \end{minipage}
    \end{subfigure}
    \begin{subfigure}[b]{.3\linewidth}
    \begin{minipage}{.1\textwidth}
        \caption{}
        \label{fig:z-topo}
    \end{minipage}%
    \begin{minipage}{0.65\textwidth}
        \includegraphics[width=\linewidth]{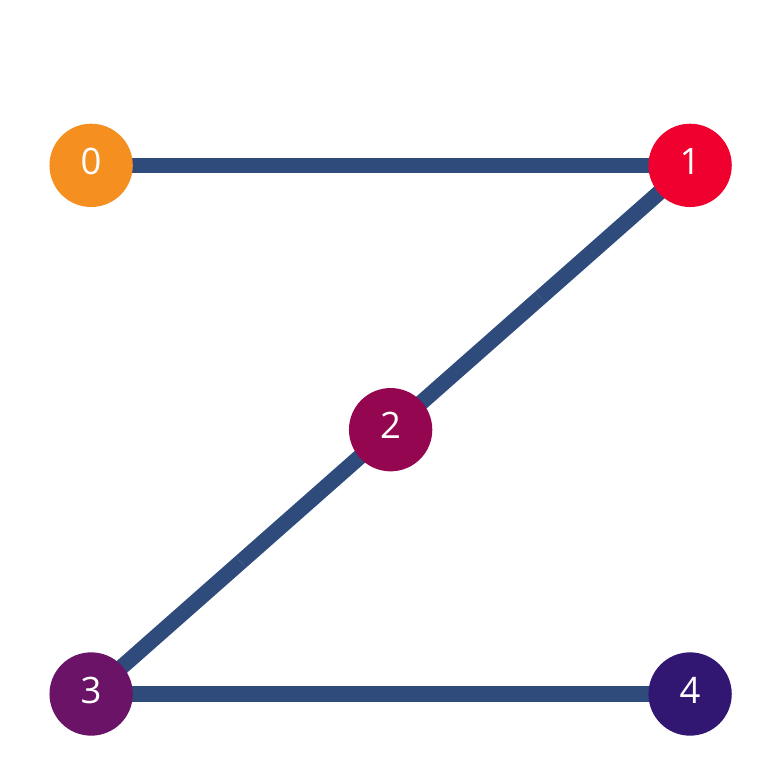}
    \end{minipage}
    \end{subfigure}
    \caption{Topologies of the IBM Quantum hardware with five qubits from which data was acquired: (a) \textit{ibmq\_rome} and \textit{imbq\_bogota}, (b) \textit{ibmq\_belem} and \textit{ibmq\_quito}, (c) \textit{ibmq\_manila}}
    \label{fig:IBMQ-systems}
\end{figure*}

\section{\label{sec:sec-3}Excited State Promoted Readout}
Excited state promoted (ESP) readout is a technique to improve qubit readout fidelity for superconducting qubits in a scalable way by using an extra level of transmon qubit, i.e., by exciting the $\ket{1}$ state to the $\ket{2}$ state for readout \cite{Jurcevic2021}. ESP is beneficial for the hardware where measurement timescales are large enough for non-negligible decay from the $\ket{1}$ state to $\ket{0}$ state. In such a system, this decay can be reduced from an extra excitation to $\ket{2}$ state, which essentially changes the discrimination problem from a two-state to a three-state system. 

\subsection{\label{subsec:sec-3-1}Theory}
In excited state promoted readout, we take advantage of higher excited states of the transmon by exciting the $\ket{1}$ state to $\ket{2}$ and then perform our measurements since $\ket{2}\to\ket{0}$ transition should be much more difficult. In order to do this, we first calibrate the frequencies amplitude of the $\pi$-pulse required for the $\ket{0}\to\ket{1}$ and the $\ket{1}\to\ket{2}$ transition using frequency spectroscopy and a Rabi experiment, respectively. Doing it for the former transition is straightforward and requires building only a gaussian wave packet. However, to assist the latter, we use a sinusoidal sideband that allows us to change the local oscillator frequency without manually setting it.

\subsection{\label{subsec:sec-3-2}Example}
In Fig. \ref{fig:IQ-plots}, we present ESP readout for \textit{ibmq\_armonk} hardware, which is an open-access Canary r1.2 one-qubit hardware from IBM. The approximate $\omega_{01}$ frequency is $4.972$ GHz, and the anharmonicity $\alpha$ is $347.19$ MHz. We show the $IQ$ plots for the $\ket{0}$-$\ket{1}$ state discrimination in Fig. \ref{fig:two-state} and the corresponding discrimination by the linear discriminant analysis (LDA). In the subsequent experiment, we use ESP and show the IQ plots in Fig. \ref{fig:three-state}. Notice that the overlap (marked by red crosses) between the $\ket{0}$ and $\ket{1}$ states decreases by exciting the $\ket{1}$ state to $\ket{2}$ state, and as a result of this increasing the confidence in readout by discriminating the correct state of the qubit.

\section{\label{sec:sec-4}Machine Learning based discriminators}
In principle, we can use machine learning methods to classify the system's different states. The first model we look at is the k-nearest neighbor (KNN) classifier, which implements learning based on the $k$ nearest neighbors of the given point for which a decision has to be taken \cite{Altman_1992}. In our case, we use $k=50$, and we calculate the "manhattan distance" to determine the proximity between the points. The second model we look at is the decision tree classifier (DTC), where the decision for the data is taken by continuously splitting it according to a certain parameter or set of rules \cite{Rokach_2013}. Our DTC model uses entropy to measure a split's quality and restrict the decision tree's depth to $20$. The third model is the classifier based on the Gaussian Naive Bayes algorithm (GNB) \cite{10.5555/1671238}. The fourth and fifth models are the linear and quadratic discriminant analysis (LDA and QDA) based classifiers \cite{Tharwat_2016}. The final model we look at is a deep learning-based model called the fully connected neural network (FNN). Our FNN architecture is composed of three hidden layers (1st, 2nd, and 3rd layer consist of $1000$, $500$, and $300$ nodes, respectively) that use ReLU activation functions, and the  $3^{N}$ output layer has \textit{softmax} activation. The network is trained (testing-validation-training set ratio of 1:1:3) using the \textit{Adam} optimizer \cite{arxiv.1412.6980} with categorical cross-entropy as the loss function.

\section{\label{sec:sec-5}Results and Discussion}
In this section, we describe the details of results for the five-qubit excited state promoted (ESP) readout experiment and compare the performances of our six models described in the previous section: (a) KNN, (b) DTC, (c) GNB, (d) LDA, (e) QDA, and (f) FNN, for the five 5-qubit IBM Quantum hardware: (i) \textit{ibmq\_rome}, (ii) \textit{imbq\_bogota}, (iii) \textit{ibmq\_belem}, (iv) \textit{ibmq\_quito}, and (v) \textit{ibmq\_manila}. 

\subsection{\label{subsec:sec-4-1}Data Accumulation and Preparation}
To perform the excited-state promoted (ESP) readout experiments, we first performed frequency calibrations and amplitude calibration experiments for each qubit $q_i$ on every given hardware to determine: $\omega_{01}$, $\omega_{12}$, $A_{01}$ and $A_{02}$, for building the $\pi_{01}$ and $\pi_{12}$ pulses. In the subsequent step, we build the pulse schedules for $3^{N}=243$ possible states, as $N = 5$ in our case. 

\begin{figure*}
    \centering
    \includegraphics[width=\textwidth]{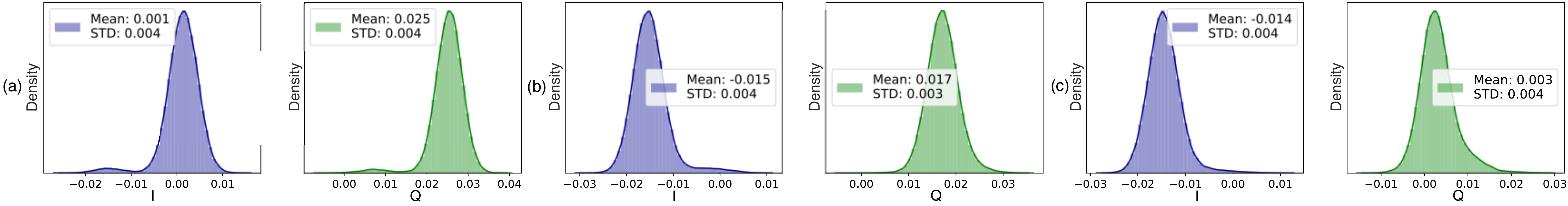}
    \caption{Probability distribution of IQ data obtained for the qubit $q_{4}$ from \textit{imbq\_belem} for states (a) $\ket{0}$, (b) $\ket{1}$ and (c) $\ket{2}$. The graphs in blue corresponds to the \textit{in-phase} component $I$ and the ones in green corresponds to the  \textit{quadrature} component $Q$.}
    \label{fig:prob}
\end{figure*}

We performed 2048 shots measurements for every such state to obtain the $IQ$ data corresponding to every state and each shot. This makes our overall data from given hardware of the size $(243,\ 2048,\ 5)$. We flatten this data over the outer dimension describing the possible number of states for preparing test, train, and validation set to be of size $(243\times2048, 10)$, where $10$ comes from splitting each of the $I+jQ$ data points into two separate values $I$ and $Q$. We then perform outlier removal using an elliptic envelope strategy \cite{Hoyle_2015} owing to the fact that individual distributions mainly follow a normalized gaussian distribution (Fig. \ref{fig:prob}). We then scale the data for individual qubits using \textit{StandardScaler} method from the \textit{sklearn} library to impose uniformity in the data points \cite{scikit-learn}. Finally, we split the data into test-train-validation sets in the proportion $(50:30:20)$. Subsequently, since we are looking at supervised learning, we also prepare the label data for training our models. While for all the machine learning models, we do so by labeling each state by the number represented by its corresponding bitstring. For example, $\ket{22102}$ would be $2\times3^{0} + 0\times3^{1} + 1\times3^{2} + 2\times3^{3} + 2\times3^{4} = 227$. Whereas, for FNN, we encode these integer labels as one-hot encoded binary vectors of size 243. 

\subsection{\label{subsec:sec-4-2}Comparison metrics}
In order to compare the performances of different discriminating models, we compute the \textit{qubit-state-assignment fidelities} $\mathcal{F}_i$, i.e., the measure of how accurately the predicted state for qubit $q_i$ matches with the correct state that it was in during measurement. For every qubit $q_i$ we define it as follows:
\begin{equation}
\begin{split}
\mathcal{F}_i = 1 - [P(0_i|\pi_{0\rightarrow 1}^i) + P(0_i|\pi_{0\rightarrow 2}^i) + P(1_i|\pi_{0\rightarrow 0}^i) \\+ P(1_i|\pi_{0\rightarrow 2}^i) + P(2_i|\pi_{0\rightarrow 0}^i) + P(2_i|\pi_{0\rightarrow 1}^i)]/6 .
\end{split}
\end{equation}
Here, we are looking at subtracting from total probability the infidelity values, i.e., deviations arising due to initialization errors, state transitions during the measurement, and readout crosstalk. In this sense, the model that can learn about these underlying causes of variations better than the others will consequently attain a better assignment fidelity score.

Furthermore, we also use another key discrimination metric $\mathcal{F}_{GM}$, that we refer to as the \textit{system-state-assignment fidelity} using the $\mathcal{F}_{i}$ for each qubit $q_i$ present on the hardware \cite{10.1103/physrevapplied.17.014024}. It is defined as the geometric mean of all the qubit-state-assignment fidelities $\mathcal{F}_{i}$:
\begin{equation}
     \mathcal{F}_{GM} = 
     (\mathcal{F}_1 \mathcal{F}_2 \mathcal{F}_3 \mathcal{F}_4 \mathcal{F}_5)^{1/5}.
\end{equation}
Finally, we also compute another metric called cross-fidelity $\mathcal{F}_{CF}$, for studying the effect of cross-talk in qubit-state assignments. For two qubits $q_i$ and $q_j$, this is defined as \cite{10.1103/physrevapplied.17.014024}:
\begin{equation}
    \mathcal{F}_{i, j}^{CF} = \mathbb{E}[1-\sum_{i, j}P(\ket{q_i}|\pi_{0\rightarrow \ket{q_{j}}\neq \ket{q_{i}}}^j)] \quad \ket{q} \in \{\ket{0},\ket{1},\ket{2}\},
\end{equation}
where $P(\ket{q_i}|\pi_{0\rightarrow \ket{q_{j}}\neq \ket{q_{i}}}^j)$ is the probability of assigning state $\ket{q_i}$ to $q_i$ when $q_j$ has been prepared in state $\ket{q_{j}} (\neq \ket{q_i}$). The (negative) positive value of $\mathcal{F}_{i,j}^{CF}$ then corresponds to the existence of (anti-) correlation between the two qubits, and for a discrimination model to handle crosstalk-induced discrimination errors effectively, its assignment correlations $\mathcal{F}_{i,j}^{CF}$ must be closer to zero for all $q_i$ and $q_j$.

\subsection{\label{subsec:sec-4-3}Key Observations}
We note some key observations from our experiments below.

\subsubsection{\label{subsec:sec-4-3-1}Single-qubit discrimination}
To measure the capabilities of each model for the single-qubit discrimination task, for each qubit $q_i$ on a given hardware, we consider the readout of the states where $\ket{q_i} \in \{\ket{0}, \ket{1}, \ket{2}\}$, and rest of the other qubits $q_{j\neq i}$ are in their ground states, i.e.,  $\ket{q_{j\neq i}} = \ket{0}^{\otimes i}\otimes\ket{q_{i}}\otimes\ket{0}^{\otimes (N-1)-i}$. Our experiments concluded that all discrimination models performed equally well after hyperparameter tuning. We attribute this behavior to the resemblance of learned boundaries for two-dimensional normalized distributions with limited overlaps (after outlier removal) by all the classifiers. This was corroborated by observing similar results in the case of standard $\ket{0}$-$\ket{1}$ readout as well.

\subsubsection{\label{subsec:sec-4-3-2}Multi-qubit discrimination}
To measure the performance for the multi-qubit discrimination task, we prepared all of the qubits in all possible $3^{N}$ states for each hardware. We then measured the individual $\mathcal{F}_{i}$ for each qubit and also the overall $\mathcal{F}_{GM}$ for each discrimination model. We present these results in the Table \ref{table:qubit-assignment-fidelity}. We see that FNN outperforms all other models for all hardware while performing equally well to some of the models (GNB and QDA) in certain qubit-state-assignment tasks. We attribute this edge to FNN's ability to learn non-linearity in the relationship between the input and output over others. 

\begin{figure}[!tp]
    \centering
    \begin{subfigure}[b]{.8\linewidth}
    \begin{minipage}{.1\textwidth}
        \caption{}
        \label{fig:gnb}
    \end{minipage}%
    \begin{minipage}{0.90\textwidth}
        \includegraphics[width=\linewidth]{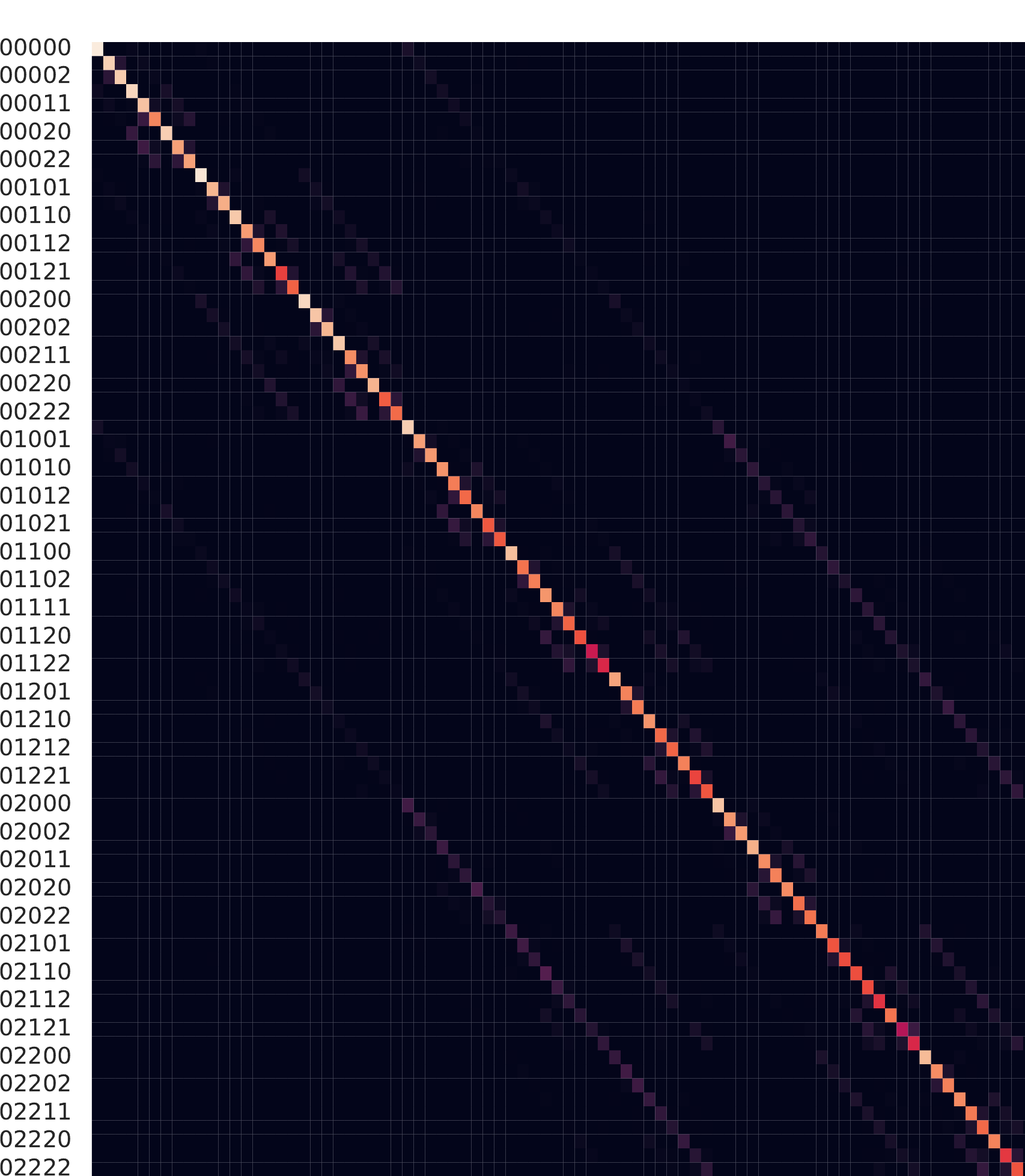}
    \end{minipage}
    \end{subfigure}
    \begin{subfigure}[b]{.8\linewidth}
    \begin{minipage}{.1\textwidth}
        \caption{}
        \label{fig:qda}
    \end{minipage}%
    \begin{minipage}{0.90\textwidth}
        \includegraphics[width=\linewidth]{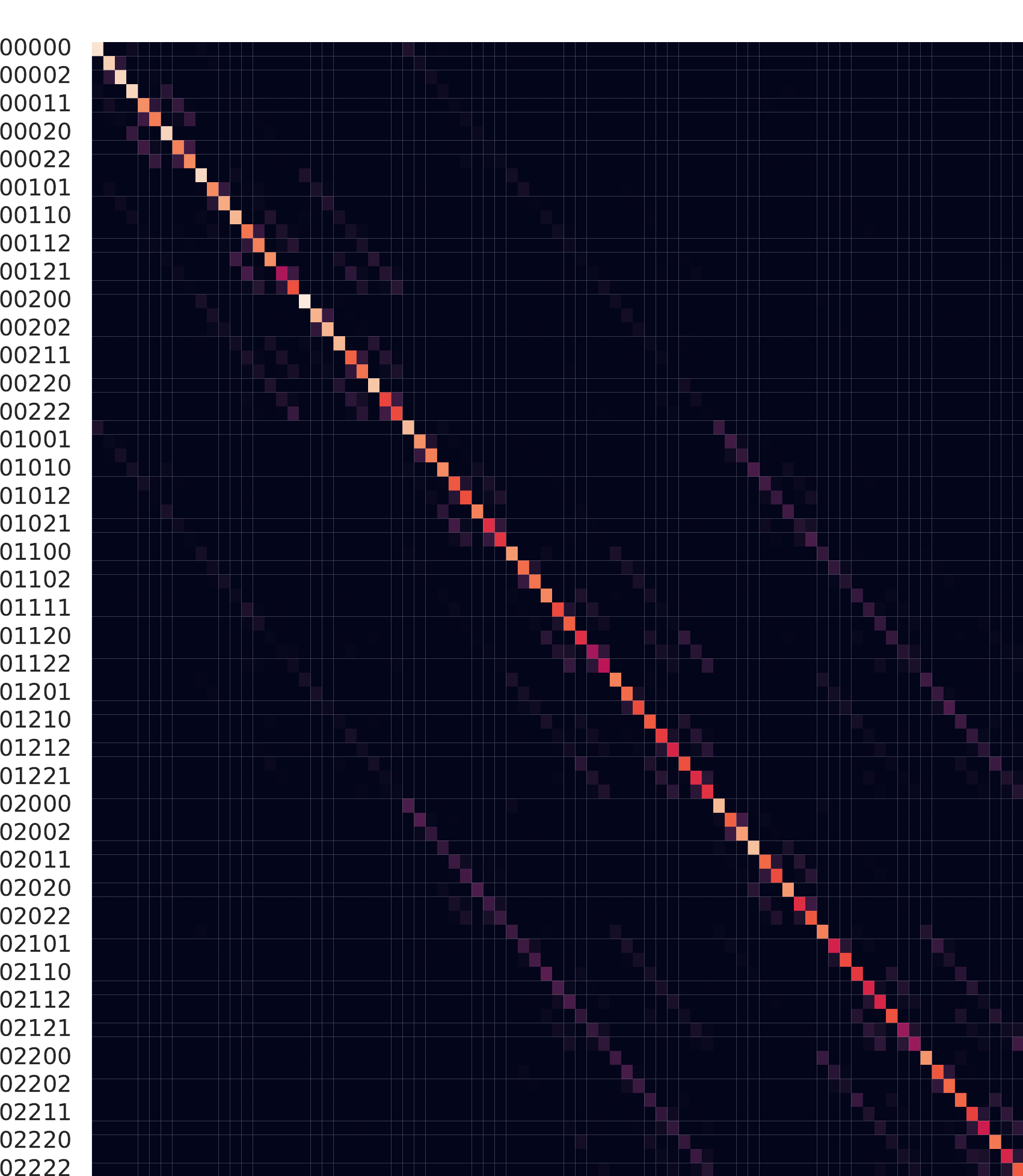}
    \end{minipage}
    \end{subfigure}\\
    \begin{subfigure}[b]{0.96\linewidth}
    \begin{minipage}{.05\textwidth}
        \caption{}
        \label{fig:fnn}
    \end{minipage}%
    \begin{minipage}{\textwidth}
        \includegraphics[width=\linewidth]{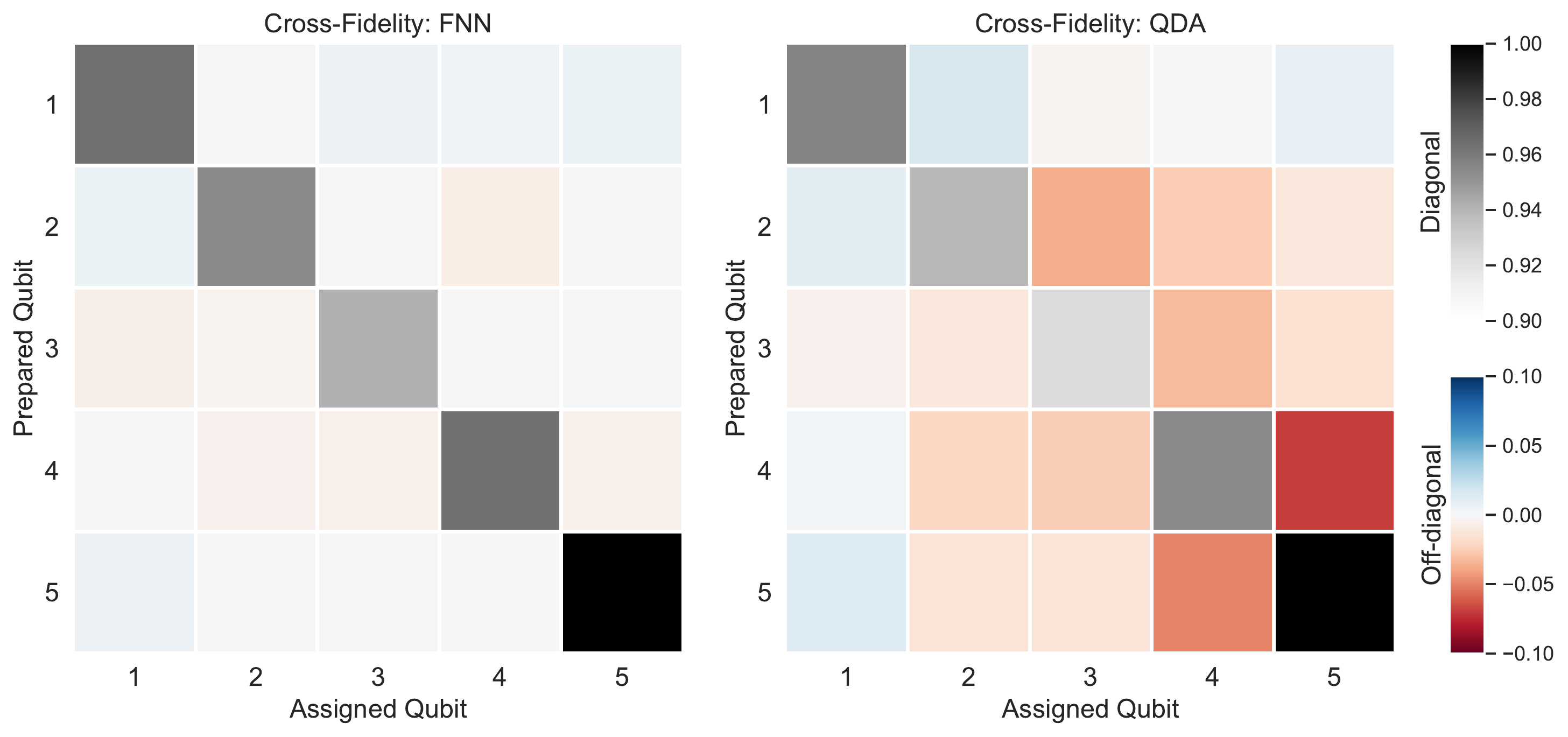}
    \end{minipage}
    \end{subfigure}
    \caption{A subset of confusion matrices for multi-qubit state assignment for \textit{ibmq\_bogota} with $\ket{q_0}=\ket{0}$ obtained using the machine learning models (a) FNN and (b) QDA. For the same models, we also show the (c) cross-fidelity matrices, where non-zero off-diagonal terms $\mathcal{F}^{CF}_{i, j}$ represent the existence of correlations between the assignment of qubit-state $\ket{q_i}$ and prepared qubit-state $\ket{q_j}$. The strength of these assignment-correlations are almost an order of magnitude smaller for FNN than for QDA for some $(q_i, q_j)$, indicating, the former's ability to learn crosstalk more effectively. For example, $\mathcal{F}_{4, 5}^{CF}$ is $-0.071$ and $-0.005$ for QDA and FNN respectively.}
    \label{fig:conf-mat}.
\end{figure}

From the data, we also see that majority of the models once again seem to perform similarly in this case well. We attribute this to either them learning the effect of crosstalk in the same way or the overlaps present in-between clusters for different states due to noise making it equally hard for all of them. To further illustrate our thinking, we look at a section of confusion and cross-fidelity matrices (for FNN and QDA) for one of the hardwares in Fig. \ref{fig:conf-mat}. We see that while most of the states misassigned by each model have similarities, FNN seems to be reducing $\mathcal{F}_{i, j}^{CF}$ by an order of magnitude (Fig. \ref{fig:fnn}).

\begin{table}[htp]
\centering

\caption{Qubits and system state-assignment fidelity comparison across hardware and classification methods. The highest fidelity values for each qubit state and overall state have been marked in bold. The final row represents order of magnitude of mean training time ($\mathcal{T}$) of each model with respect to $\mathcal{T}_{\text{GNB}}$.}
\begin{tabular}{clcccccc}
\textbf{Hardware} &                                          & \textbf{KNN} & \textbf{DTC} & \textbf{GNB} & \textbf{QDA} & \textbf{LDA} & \textbf{FNN}  \\ 
\hline
& & & & & & & \\
                  & $\mathcal{F}_{1}$  & 0.913        & 0.891        & 0.918        & 0.919        & 0.918        & \textbf{0.923}         \\
                  & $\mathcal{F}_{2}$  & 0.918        & 0.914        & 0.925        & \textbf{0.927}        & 0.926        & \textbf{0.927}         \\
ibmq\_rome              & $\mathcal{F}_{3}$  & 0.975        & 0.963        & 0.978        & 0.978        & 0.978        & \textbf{0.981}         \\
                  & $\mathcal{F}_{4}$  & 0.940        & 0.936        & 0.943        & 0.945        & 0.943        & \textbf{0.949}         \\
                  & $\mathcal{F}_{5}$  & 0.924        & 0.910        & 0.930        & 0.932        & 0.930        & \textbf{0.938}         \\
                  & $\mathcal{F}_{GM}$ & 0.934        & 0.933        & 0.939        & 0.939        & 0.939        & \textbf{0.943}         \\ 

\hline
\\
                  & $\mathcal{F}_{1}$  & 0.927        & 0.943        & 0.943        & 0.943        & 0.940        & \textbf{0.947}         \\
                  & $\mathcal{F}_{2}$  & 0.941        & 0.953        & 0.953        & 0.953        & 0.951        & \textbf{0.957}         \\
ibmq\_bogota      & $\mathcal{F}_{3}$  & 0.969        & 0.975        & 0.975        & 0.973        & 0.975        & \textbf{0.981}         \\
                  & $\mathcal{F}_{4}$  & 0.980        & 0.984        & 0.983        & 0.983        & 0.983        & \textbf{0.989}         \\
                  & $\mathcal{F}_{5}$  & 0.896        & 0.916        & 0.915        & 0.914        & 0.910        & \textbf{0.923}         \\
                  & $\mathcal{F}_{GM}$ & 0.937        & 0.890        & 0.943        & 0.946        & 0.946        & \textbf{0.960}         \\ 
\hline
\\
                  & $\mathcal{F}_{1}$  & 0.957        & 0.946        & 0.958        & 0.958        & 0.958        & \textbf{0.963}         \\
                  & $\mathcal{F}_{2}$  & 0.950        & 0.936        & 0.950        & 0.950        & 0.950        & \textbf{0.964}         \\
ibmq\_belem             & $\mathcal{F}_{3}$  & 0.902        & 0.882        & \textbf{0.904}        & 0.903        & 0.902        & \textbf{0.904}         \\
                  & $\mathcal{F}_{4}$  & 0.987        & 0.983        & 0.988        & 0.988        & 0.987        & \textbf{0.993}         \\
                  & $\mathcal{F}_{5}$  & 0.977        & 0.971        & 0.979        & 0.979        & 0.978        & \textbf{0.987}         \\
                  & $\mathcal{F}_{GM}$ & 0.954        & 0.943        & 0.955        & 0.955        & 0.955        & \textbf{0.962}         \\ 
                  
\hline
\\
                  & $\mathcal{F}_{1}$  & 0.942        & 0.927        & 0.943        & 0.943        & 0.943        & \textbf{0.959}         \\
                  & $\mathcal{F}_{2}$  & 0.952        & 0.941        & 0.953        & 0.953        & 0.953        & \textbf{0.961}         \\
ibmq\_quito             & $\mathcal{F}_{3}$  & 0.974        & 0.969        & 0.975        & 0.975        & 0.973        & \textbf{0.983}         \\
                  & $\mathcal{F}_{4}$  & 0.982        & 0.980        & 0.984        & 0.983        & 0.983        & \textbf{0.992}         \\
                  & $\mathcal{F}_{5}$  & 0.912        & 0.896        & 0.916        & 0.915        & 0.914        & \textbf{0.931}         \\
                  & $\mathcal{F}_{GM}$ & 0.952        & 0.942        & 0.954        & 0.954        & 0.953        & \textbf{0.965}         \\ 

\hline
\\
                  & $\mathcal{F}_{1}$  & 0.922        & 0.904        & 0.926        & 0.925        & 0.926        & \textbf{0.937}         \\
                  & $\mathcal{F}_{2}$  & 0.915        & 0.895        & 0.919        & 0.918        & 0.919        & \textbf{0.921}         \\
ibmq\_manila            & $\mathcal{F}_{3}$  & 0.908        & 0.887        & \textbf{0.913}        & 0.912        & \textbf{0.913}        & \textbf{0.913}         \\
                  & $\mathcal{F}_{4}$  & 0.922        & 0.903        & 0.925        & 0.924        & 0.925        & \textbf{0.932}         \\
                  & $\mathcal{F}_{5}$  & 0.944        & 0.931        & 0.948        & 0.947        & 0.948        & \textbf{0.961}         \\
                  & $\mathcal{F}_{GM}$ & 0.922        & 0.904        & 0.926        & 0.925        & 0.926        & \textbf{0.933}         \\

\hline \\
\multicolumn{2}{c}{$\log_{10}(\mathcal{T}/\mathcal{T}_{\text{GNB}})$} & 0.723        & 3.761        & \textbf{0.0}        & 0.103        & 0.534        & 1.241         \\
\label{table:qubit-assignment-fidelity}
\vspace{-20pt}
\end{tabular}
\end{table}

\subsubsection{\label{subsec:sec-4-3-3}Accuracy vs. training time tradeoff}
We fit (or trained) all of our models for an explicitly similar amount of data points $~(243\times1024)$. While DTC was easily the most expensive model regarding average training time ($\mathcal{T}$), the GNB came out to be quickest, and that too with, in general, good qubit-state and system-state assignment-fidelities (Table \ref{table:qubit-assignment-fidelity}). For FNN, we limited the training epochs to a fixed number as we saw that the network was beginning to overfit due to the presence of large density layers while being trained for extended periods. One way to overcome this overfitting would be to include dropout layers in addition to the dense layers. Still, at the same time, that might lead to the loss of connections between different data attributes that might have been essential for effective multi-class classification. Hence, for this reason, the average training time of FNN turned out to be competitive with the rest of the models but still a lot slower than that of GNB ($\log_{10}(\mathcal{T}_{\text{FNN}}/\mathcal{T}_{\text{GNB}}) = 1.241$).

\section{Conclusion}

We have compared various machine learning-based discrimination models for multi-qubit readout tasks. We found FNN to be more crosstalk-resilient than other approaches. At the same time, GNB is more efficient if we consider the accuracy-training time tradeoff and the training data size robustness. This makes the latter a good choice compared to FNN in cases where training time latency is crucial, and the size of training samples is limited. 
While the authors in \cite{10.1103/physrevapplied.17.014024} have previously noted similar performance improvements when using FNN for two-state discrimination, their analysis was restricted to a limited subset of discrimination models studied here and that too for just one five-qubit hardware.
In the future, we would like to focus on the task of unsupervised clustering \cite{Alashwal_2019} for qubit readout using techniques like Gaussian mixture models \cite{Guoshen_Yu_2012} owing to the normalized distribution of the $IQ$ data points and also looking at their robustness to noise and crosstalk effects with the scaling of the quantum hardware. 


\section*{Acknowledgement}
This work has been done as part of the Qiskit Advocate Mentorship Program (QAMP), and we thank its organizers for the same. 
The views expressed are those of the authors and do not reflect the official policy or position of IBM or the IBM Quantum team.

\bibliographystyle{ieeetr}
\bibliography{mlbdespr}

\end{document}